\documentclass[conference]{IEEEtran}
\IEEEoverridecommandlockouts
\usepackage{cite}
\usepackage{amsmath,amssymb,amsfonts}
\usepackage{algorithmic}
\usepackage{graphicx}
\usepackage{textcomp}
\usepackage[letterpaper,
             left=0.625in,
             right=0.625in,
            top=0.75in,
         bottom=1in,]{geometry}             

\usepackage{tablefootnote}

\usepackage{multirow}
\usepackage{multicol}
\usepackage{soul,xcolor}
\usepackage[normalem]{ulem}

\newcommand{\task}{\langle j, t \rangle}
\newcommand{\taskset}{\langle j, t \rangle \in \langle \mathcal{J}, \mathcal{T} \rangle}
\graphicspath{{figures/}}

\graphicspath{{figures/}} 

\begin{document}

%\title{Risk Evaluating Deep Reinforcement Learning for Task Offloading in Smart Farm Networks}
\title{To Risk or Not to Risk: Learning with Risk Quantification for IoT Task Offloading in UAVs}
\author{\IEEEauthorblockN{Anne Catherine Nguyen$\dag$, Turgay Pamuklu$\dag$, \IEEEmembership{Member, IEEE}, Aisha Syed$\ddag$, \\ W. Sean Kennedy$\ddag$, Melike Erol-Kantarci$\dag$, \IEEEmembership{Senior Member, IEEE}}

\IEEEauthorblockA{$\dag$\textit{School of Electrical Engineering and Computer Science,}
\textit{University of Ottawa}, Ottawa, Canada}

\IEEEauthorblockA{$\ddag$\textit{Nokia Bell Labs}\\
Emails:\{anguy087, turgay.pamuklu, melike.erolkantarci\}@uottawa.ca, 
\{aisha.syed, sean.kennedy\}@nokia-bell-labs.com}
}

\maketitle
\makeatletter
\def\ps@IEEEtitlepagestyle{%
  \def\@oddfoot{\mycopyrightnotice}%
  \def\@oddhead{\hbox{}\@IEEEheaderstyle\leftmark\hfil\thepage}\relax
  \def\@evenhead{\@IEEEheaderstyle\thepage\hfil\leftmark\hbox{}}\relax
  \def\@evenfoot{}%
}

\def\mycopyrightnotice{
 \begin{minipage}{\textwidth}
 \centering \scriptsize
Accepted for ICC2023. © 20XX IEEE.  Personal use of this material is permitted.  Permission from IEEE must be obtained for all other uses, in any current or future media, including reprinting/republishing this material for advertising or promotional purposes, creating new collective works, for resale or redistribution to servers or lists, or reuse of any copyrighted component of this work in other works.
 \end{minipage}
}
\makeatother

\begin{abstract}
A deep reinforcement learning technique is presented for task offloading decision-making algorithms for a multi-access edge computing (MEC) assisted unmanned aerial vehicle (UAV) network in a smart farm Internet of Things (IoT) environment. The task offloading technique uses financial concepts such as cost functions and conditional variable at risk (CVaR) in order to quantify the damage that may be caused by each risky action. The approach was able to quantify potential risks to train the reinforcement learning agent to avoid risky behaviors that will lead to irreversible consequences for the farm. Such consequences include an undetected fire, pest infestation, or a UAV being unusable. The proposed CVaR-based technique was compared to other deep reinforcement learning techniques and two fixed rule-based techniques. The simulation results show that the CVaR-based risk quantifying method eliminated the most dangerous risk, which was exceeding the deadline for a fire detection task. As a result, it reduced the total number of deadline violations with a negligible increase in energy consumption.    
\end{abstract}

\begin{IEEEkeywords}
Risk quantification, Unmanned Aerial Vehicles,  Deep Reinforcement Learning, Smart Farm
\end{IEEEkeywords}
\section{Introduction}
Wireless technology has expanded the horizon of the agriculture industry. It has enabled more efficient and precise farming techniques with the deployment of the Internet of Things (IoT) devices and wireless connectivity. Farmers are now able to monitor their farmlands using sensors and cameras that relay back to  their real-time status updates. Through image classification, IoT devices can quickly identify changes in environments and potential risks. Qazi et al. \cite{qazi2022iot} highlighted the latest advancements in IoT technologies and artificial intelligence used in smart farms. The authors identified that with the assistance of artificial intelligence, smart farm networks are able to self-manage their resources in order to maintain efficient performance.
 
Furthermore, energy-aware reinforcement learning solutions are recently proposed for many wireless network problems. Mollahasani et al. \cite{Mollahasani2022} proposed actor-critic-based learning to reduce energy consumption in an open radio access network architecture. Khoramnejad et al. \cite{Khoramnejad2022} addressed energy consumption and quality of service performance metrics in 5G networks by using a multi-agent double deep Q-network. Pamuklu et al. \cite{Pamuklu2021} included solar panels in their architecture as an alternative energy source for reducing the dependency on grid energy. Their reinforcement learning solution also improved the cost efficiency to improve the economic feasibility. Energy-aware-based machine learning (ML) solutions are also attracting interest in unmanned aerial vehicle (UAV) based wireless networks recently \cite{Abrar2021}. Sun et al. \cite{Sun2021} modeled their age of information and energy-centric problem as a Markov decision process and then solved it with a policy gradient-based machine learning algorithm.

Task offloading is another emerging topic in UAV-based networks. Ebrahim et al. \cite{Ebrahim2022} provided a tradeoff between delay and energy by optimizing offloading decisions using deep reinforcement learning (RL). Sacco et al. \cite{Sacco2022} aimed to reduce the required information and training time for an RL-based task offloading solution. Zhao et al. \cite{Zhao2022} studied on multi-agent TD3 approach to address their continuous action space problem. Yang et al. \cite{Yang2022} focused on service cost minimization by optimizing the joint task offloading and time-division multiple access-based channel allocation decisions.

In our previous works, we also studied task offloading problems in UAV-based smart agriculture IoT networks. First, we proposed a tabular Q-Learning approach \cite{Nguyen2022a}, and then a deep RL version \cite{Nguyen2022} for larger networks. In addition to these studies, we proposed a risk-sensitive solution \cite{Pamuklu2022} to isolate critical key performance indicators; thus, we can control the tradeoff between multi objectives while training our tabular ML model.

Unlike our prior works, we also looked into risk quantification techniques. In finance, there are several methods used to measure risks, such as the Sharpe ratio, the Sortino ratio, and conditional value at risk (CVaR). These measurements are used to evaluate the riskiness of investments based on their expected losses \cite{rockafellar2000optimization}. The Sharpe ratio \cite{sharpe1994} measures the risk of an investment portfolio by dividing the increase in return by the volatility of the portfolio. First, it subtracts the return of a known safe investment from the portfolio's expected return. Then the difference between the two portfolios is divided by the standard deviation of the first portfolio's returns. The difference between the two portfolios indicates the potential increase in return, and the standard deviation of the investment's return indicates the volatility of the returns. The Sortino ratio \cite{sortinoprice1994} is an extension of the Sharpe ratio, but it has a different denominator. Instead of dividing by the standard deviation of all the portfolio's returns, it uses the standard deviation of the portfolio's negative returns. CVaR was introduced by Rockafellar and Uryasev as a way to evaluate investment portfolios based on their expected shortfalls \cite{rockafellar2000optimization}. CVaR is the mean of the lowest $\alpha$ percentile of the portfolio's returns. This enables investors to say that the portfolio's expected returns are better than the CVaR $(1-\alpha)$ percent of the time.

In wireless networks, financial concepts like risk can be used to evaluate and model behaviors to find an algorithm that avoids undesirable behaviors. Apandi et al. \cite{Apandi} use the Sharpe ratio as a decision-making parameter for base stations to use in order to find an optimal way to find user associations in order to send subcarriers. Zhou et al. \cite{ZhouAoI} use CVaR to measure the average of the worst age of information times for IoT status updates at the receiver in an IoT monitoring system. The CVaR of the receiver's age of information was used as one of the considerations in their optimization algorithm. Unlike this paper, we use CVaR to measure the potential damages after certain dangerous events happen in a smart farm. 

The rest of this paper is organized as follows: Sections II and III describe the system model and the proposed CVaR-based risk quantifying approach, respectively. Then, we present the performance evaluation of this method in Section IV and conclude the paper with Section V.
\section{System Model}
\begin{figure}
    \centering
    \includegraphics[width=0.9\linewidth]{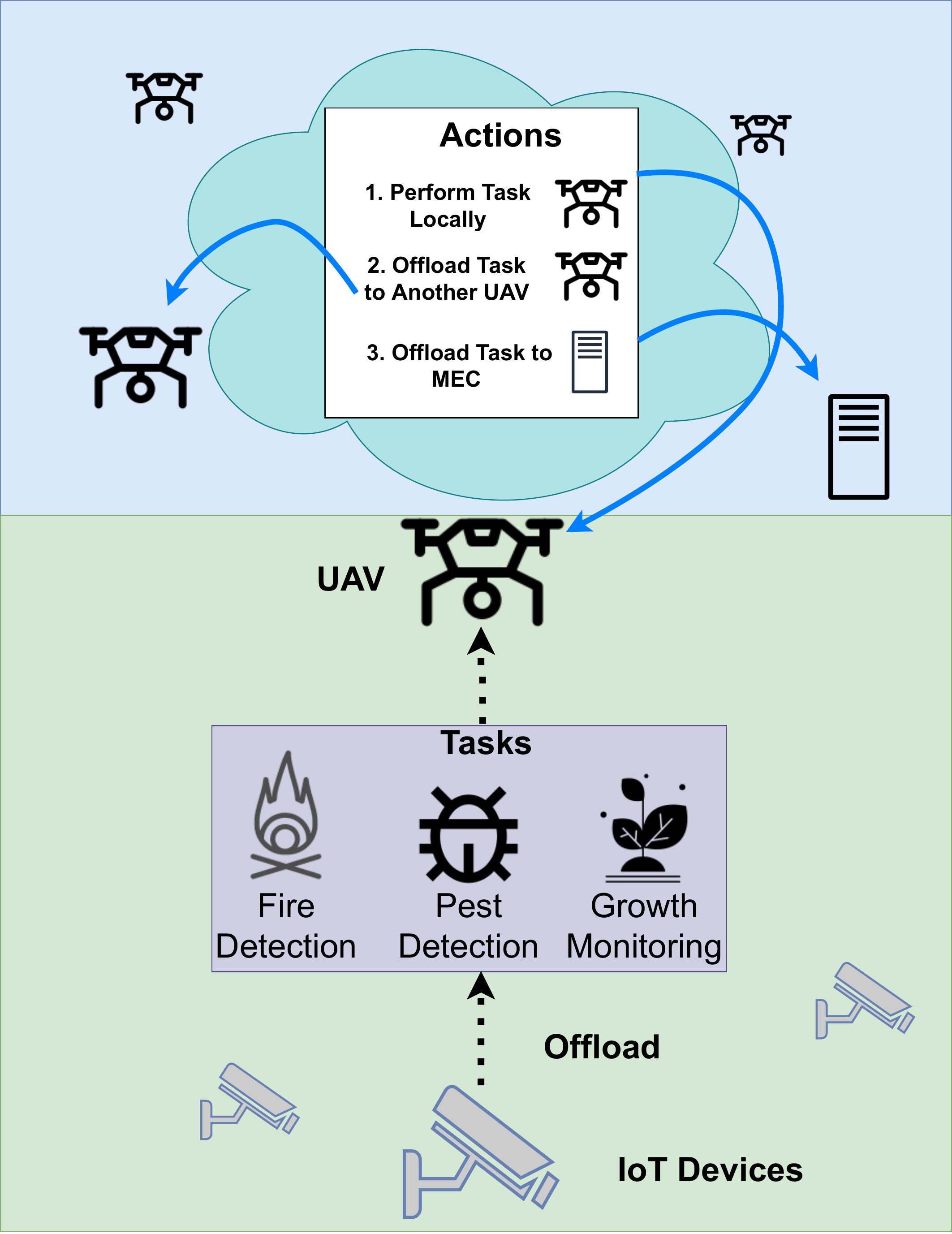}
    \caption{\label{fig:system_model} The network model for risk quantifying approach.}
\end{figure}
Our smart farm scenario consists of a set of $X$ IoT devices deployed across large farmland. These IoT devices perform real-time intensive image classification tasks in order to provide up-to-date updates on the farm. They can perform $K$ different types of tasks such as fire detection, pest detection, and crop growth monitoring. One of the limitations of IoT devices being used on farms is their finite battery and computing capacity. Accurate image classification tasks require rigorous computation because they need to use deep neural networks (DNNs). If the IoT devices perform all the image classification tasks themselves, they will run out of batteries very quickly while trying to meet the heavy computing resource demands of the image classification algorithms. In addition, some of these tasks need to be done within a certain time frame; therefore they have deadlines. 

In order to relieve some of the demands of the IoT devices, we have $J$ unmanned aerial vehicles (UAV) deployed over the farmlands. A UAV device ($j$) contains a computing resource that is capable of performing an image classification task $\task$, locally, which is generated in time interval $t$. Also, a UAV provides connectivity to other UAVs or multi-access edge computing (MEC) devices ($\mathcal{L}$) where this task can be performed. With the addition of these computing resources ($J^{+}=\mathcal{J}\cup\mathcal{L})$, IoT devices can offload their tasks to nearby UAVs. Upon receiving a task, the UAVs can choose to compute the task themselves, offload to another UAV, or offload to a MEC server. The UAVs also suffer from having a finite battery capacity $\Upsilon_{j'}^{B}$. Their current battery level at time $\mathcal{T}$, ($\Upsilon_{j'}^{R}$) can be modeled using the following equation,
\begin{flalign}
    \label{eq:energyeq}
    & \Upsilon^{R}_{j'} = \Upsilon^{B}_{j'} - (\Upsilon^{H}_{j'} + \Upsilon^{A}_{j'} + \Upsilon^{I}_{j'}) * \mathcal{T} \notag \\ 
    &- \sum\limits_{\substack{j\in\mathcal{J}\\ t\in\mathcal{T} \\ t'\in\mathcal{T}}} (\Upsilon^{C}_{j'} - \Upsilon^{I}_{j'})*p_{\task j't'},
\end{flalign}
where $\Upsilon^{B}_{j'}$ is the initial battery level  for UAV $j'$, $\Upsilon^{H}_{j'}$ is the amount of energy required for the UAV to fly above the farm, $\Upsilon^{A}_{j'}$ is the amount of energy required by the antenna to send signals, $\mathcal{T}$ is the total simulation time, $\Upsilon^{I}_{j'}$ is the amount of energy required for the CPU to be idle, and $\Upsilon^{C}_{j'}$ is the amount of energy required for the CPU to run a task. $p_{\task j't'}$ is a binary indicator used to indicate that UAV $j'$ computed the task $\task$ at time $t'$. Computing a task locally will lead to some of their battery power being consumed; however, if the current UAV has too many tasks to compute, the task's queuing time will increase, which can lead to a deadline violation occurring $\mathbb{E}(v_{\task}) = 1$ because the task's processing time exceeds the task's deadline. 

One of the objectives of the proposed methods is to extend the UAV's battery life. This will enable the UAV to hover for as long as possible. The next objective is to minimize the average mean uplink delay for task completion. By the uplink delay, we are reducing the chances of a deadline violation occurring. Not meeting the tasks' deadline $\gamma^{D}_{\task}$ can have severe consequences. For example, if the pests are not detected in time, a significant portion of the crops may be destroyed. These consequences can be described as risky behavior that we would like to avoid. We are proposing two methods that will make decisions on behalf of the UAVs while considering the consequences of the risks during the decision-making process.      

\section{Risk Quantifying (RQ) Method}
\subsection{Deep Risk Sensitive Learning with Risk Measurement}
Deep reinforcement learning uses deep neural networks to generate Q-values for each action instead of Q-tables. These Q-values indicate to the agent the action that will lead to the best long-term gain. The agent then selects the action. As the agent goes through a series of actions and encounters different scenarios, the neural networks train themselves with new experiences to get more precise in predicting the Q-values. In addition, the neural networks allow a larger state space without being bound by physical disk space like the tabular Q-learning method. This enabled us to include more environmental information in the state such as the transmission delays between all of the computing resources. When dealing with a multi-objective problem, we need to juggle many risks at the same time and evaluate each risk. We need a measure of risk in order to properly quantify each risk so that we can determine the correct course of action. In the field of finance, there are several methods of evaluating risk in order to select the best investment. Our approach to evaluating the performance of each objective is based on the CVaR calculation. 

The CVaR method is particularly applicable to our smart farm scenario because we have several different types of risks that can occur at the same time throughout monitoring the farm. Each risk can lead to different levels of damage. For example, when you miss the deadline of a fire recognition task by one second, it will cause some containable damage. The longer you exceed the deadline, the amount of damage will increase drastically and become unmanageable. Similarly, when the UAV is operating at full battery capacity, performing a task locally does not do a lot of damage since it will only consume a small percentage of the battery power. However, when the UAV's battery is almost depleted, performing a task locally now becomes more dangerous because it may cause the UAV to stop working. The potential damage of each known risk factor can be modeled according to the objectives listed in (\ref{eq:objCVAR}) as a function of time, and the sum of all the risk factors can be viewed similarly to how we view the potential returns of an investment portfolio. Using CVaR will allow us to train the agent to avoid actions that will lead to the worst damages.  

Instead of only considering the returns' lowest $\alpha$ percentile, we consider the mean of both the highest and lowest $\alpha/2$-percentile. We found that if you only use CVaR, you are only avoiding the worst behavior. The agent only learns not to do actions that lead to the worst damages; however, it will still perform actions that lead to damages that are slightly better than the worst damages. When you include the upper $\alpha/2$-percentile in the evaluation, you can train the agent to learn good behavior along with avoiding bad ones. 

Unlike the deep risk-sensitive approach presented in Section \ref{sec:DRS}, this approach does not have a separate state to evaluate the known risk. Also, it does not consider all deadline violations to be the same. It uses cost functions to evaluate the value of each risk based on the potential damages. 

\subsubsection{Objective}
The objectives of this problem are similar to the objectives highlighted in \cite{Pamuklu2022} where we would like to minimize the hovering time of the UAV network ($\Upsilon^{R}_{j'}$) , and minimize the mean uplink delay ($\delta$). Instead of restricting the total number of deadline violations, we would like to minimize the total number of deadline violations that will lead to severe consequences ${E}(v_{\task}) = 1$. For example, if we exceed the deadline for a fire identification task, the consequences from this deadline violation will lead to more damage than if the deadline for a growth monitoring task is exceeded. \\
\textbf{Maximize:}
\begin{flalign}
    \label{eq:objCVAR}
    W * \min_{ j' \in J} \Upsilon^{R}_{j'} - \frac{1-W}{2*\Theta^{M}} \delta  
    &- \frac{1-W}{2*\Theta^{D}} \sum\limits_{\taskset}\mathbb{E}(v_{\task})
\end{flalign}

\subsubsection{Risk Cost Functions}
The cost functions determine the potential cost of damages an action can inflict on the smart farm. The costs for each risk is evaluated after every action, and the sum of all the individual cost forms the total cost. 
\begin{itemize}
    \item Energy Risk Cost ($\mathcal{C}_{e}$): As the UAV's current battery power after the action ($\Upsilon^{R}_{j_{a}}$) goes down, the situation becomes more serious. This is reflected in exponential growth ($g$) seen in (\ref{eq:energycostfunc}).
    \begin{flalign}
        \label{eq:energycostfunc}
        &\mathcal{C}_{e}(\Upsilon^{R}_{j_{a}}) = (1-\Upsilon^{R}_{j_{a}})^{g}
    \end{flalign}

    \item End-to-end Delay Risk Cost ($\mathcal{C}_{d}$): The end-to-end delay risk is a function of the task's end-to-end delay ($\Delta_{\task}$). The end-to-end delay can be defined as 
    \begin{flalign}
        \label{eq:endtoenddelay}
        &\Delta_{\task} = \Delta_{I}^{j_{R}} + \Delta_{j_{R}}^{j_{P}} + \Upsilon^{Q}_{\task} + \Upsilon_{\task}^{P},
    \end{flalign}
    where $\Delta_{I}^{j_{R}}$ is the transmission delay from the IoT device ($I$) to the UAV that first received the task ($j_{R}$), $\Delta_{j_{R}}^{j_{P}}$ is the transmission delay from $j_{R}$ to the computing resource that will compute the task $j_{P}$, $\Upsilon^{Q}_{\task}$ is the time the task spends in $j_{P}$'s queue, and $\Upsilon_{\task}^{P}$ is the time it takes for the task to be computed at by $j_{P}$.
    The ideal situation is when the end-to-end delay is less than the task's deadline ($\gamma^{D}_{\task}$). This is why it will have the lowest cost; the more the end-to-end delay is lower than the deadline. If the end-to-end delay is equal to the deadline, it is still acceptable, which is why the $\mathcal{C}_{d}$ is 0. If the end-to-end delay is greater than the task's deadline, then the cost will be positive because there will be some damage done to the smart farm environment. The magnitude of  the cost $E(\Delta_{\task})$ depends on the end-to-end delay, the deadline, and the task type. If the task's type is a fire task ($h_{\task} = 1$), the costs associated with missing this task's deadline will be more severe and, therefore, higher. The further away the end-to-end delay is from the task's deadline, the magnitude of the cost will increase by a power of $s$. If the task type is not a fire task ($h_{\task} = 0$), the magnitude of the cost will increase by a power of $w$. 
    \begin{flalign}
        \label{eq:DVcostfunc}
        &\mathcal{C}_{d}(\Delta_{\task}) = 
        \begin{cases}
            -E(\Delta_{\task}) \text{ if } \Delta_{\task} < \gamma^{D}_{\task} \\
            0 \text{ if } \Delta_{\task} = \gamma^{D}_{\task} \\ 
             E(\Delta_{\task}) \text{ if } \Delta_{\task} > \gamma^{D}_{\task}
        \end{cases}  \\ \notag
        \label{eq:DVcostfuncpt2}
        &\textbf{Where: } E(\Delta_{\task}) = h_{\task}|(\gamma^{D}_{\task}-\Delta_{\task})|^s \\
        & + (1-h_{\task})|(\gamma^{D}_{\task}-\Delta_{\task})|^w 
    \end{flalign}

    \item Total Cost ($\mathcal{C}_{T}$): Total cost is the sum of the energy risk cost ($\mathcal{C}_{e}$) and the end-to-end delay cost ($\mathcal{C}_{d}$). $\Gamma$ is the scaling factor for the energy cost.
    \begin{flalign}
        \label{eq:totalcost}
        &\mathcal{C}_{T} = \Gamma * \mathcal{C}_{e} + \mathcal{C}_{d}
    \end{flalign}
\end{itemize}

\subsubsection{Risk Measurement}
After every action, the total cost of that action ($\mathcal{C}_{T_{a}}$) is calculated and we form an ordered list of costs ($\mathbb{Z}_{j_{a}}$) for all actions performed by agent $j_{a}$. We assume that agent $j_{a}$ has performed a total of $n$ actions at time $t$ (in seconds). Let $\mathbb{Z}_{L_{j_{a}}}$ be a list that contains the lowest $\frac{\alpha}{2} $\% of $n$ elements in  $\mathbb{Z}_{j_{a}}$. Let $\mathbb{Z}_{H_{j_{a}}}$ be a list that contains the highest $\frac{\alpha}{2} $\% of $n$ elements in $\mathbb{Z}_{j_{a}}$. The risk measurement ($\eta$) for agent $j_{a}$ after action $a$ is defined as 

\begin{flalign}
    \label{eq:riskmeasurement}
    &\eta = \frac{\sum \mathbb{Z}_{L_{j_{a}}} + \sum \mathbb{Z}_{L_{j_{a}}}}{\frac{\alpha}{100} * n}. 
\end{flalign}

\subsubsection{Markov Decision Process Definition} 
\begin{itemize}
    \item \textbf{State: } The state includes the task type $k$, the CPU delays of all computing resources in UAVs and MECs $\Delta_{j'\in\mathcal{J^{+}}}$, all UAVs' current battery levels $\Upsilon^{L}_{j'\in\mathcal{J}}$, all transmission delays between computing resources $\Delta^{j_P \in\mathcal{J^+}}_{j_R \in\mathcal{J^+}}$, and the UAV that first received the task $j_i$. The state can be defined as
    \begin{flalign}
        \label{eq:state}
        &\mathbb{S}= \{k, \Delta_{j'\in\mathcal{J^{+}}}, 
        \Upsilon^{L}_{j'\in\mathcal{J}}, 
        \Delta^{j_P \in\mathcal{J^+}}_{j_R \in\mathcal{J^+}}, 
        j_i\}.
    \end{flalign}
    \item \textbf{Action: } Each UAV will act as an independent agent. Upon receiving a task, an agent must select a destination $j'$ which is a computing resource, out of all the computing resources available $J^{+}$, where the task will be computed. The agent can choose to do the task on the local computing resource or send the task to a nearby UAV or MEC server. Therefore, the set of all possible actions is defined as
    \begin{flalign}
        \label{eq:action}
        \mathbb{A} = \{ x_{j' \in J^{+}} \}.
    \end{flalign}

\end{itemize}

\par The reward function $\mathbb{R}_{RQ}$ is defined as 
\begin{flalign}
    \label{eq:rewCVAR}
    &\mathbb{R}_{RQ} = (1-\beta) RV_{c} + \beta RV_{m},
\end{flalign}             
where $RV_{c}$ is the reward value of the current UAV, $RV_{m}$ is the reward value of the UAV with the highest costs, and $\beta$ is the degree to which we are considering the UAV with the highest costs. The values for $RV_{m}$ and $RV_{c}$ are defined using (\ref{eq:cvarrv}), and depend on the UAV's CVaR value before ($\eta_{p}$) and after ($\eta_{a}$) the action was taken. If $\eta_{a} < \eta_{p}$, then the action caused the average risk cost to decrease, and the agent will be rewarded with the highest reward. If $\eta_{a} < \eta_{p}$, then the action caused the average risk cost to increase which means that the agent is engaging in actions that are significantly increasing battery consumption, uplink mean delay, or exceeding the deadline violation of a dangerous task. Therefore, the reward value is negative.
\begin{flalign}
    \label{eq:cvarrv} &
    RV = 
    \begin{cases}
            20, \text{ if }(\eta_{a} < \eta_{p}) \land (\eta_{a} < 0) \\
            10, \text{ if }(\eta_{a} < \eta_{p}) \land (\eta_{a} = 0) \\
            1, \text{ if }(\eta_{a} < \eta_{p}) \land (\eta_{a} > 0) \\
            5, \text{ if }(\eta_{a} = \eta_{p}) \land (\eta_{a} < 0) \\
            2, \text{ if }(\eta_{a} = \eta_{p}) \land (\eta_{a} = 0) \\
            -5, \text{ if }(\eta_{a} = \eta_{p}) \land (\eta_{a} > 0) \\
            1, \text{ if }(\eta_{a} > \eta_{p}) \land (\eta_{a} < 0) \\
            -1, \text{ if }(\eta_{a} > \eta_{p}) \land (\eta_{a} = 0) \\
            -10, \text{ if }(\eta_{a} > \eta_{p}) \land (\eta_{a} > 0) \\
    \end{cases}
\end{flalign}

\subsection{Baseline Methods}
\par \subsubsection{Round Robin (RR)}
In the round robin, there is a fixed order to determine the destination of the task. All computing resources are placed on an ordered list, and the task's destination is determined by knowing the previous task's destination and sending the current task to the next destination on the ordered list.

\par \subsubsection{Lowest Queue Time and Highest Energy First (QHEF)} 
Similar to the RR algorithm, QHEF uses fixed rules to determine the destination of the task. The algorithm determines the destination where the task will be computed by first finding the destination with the lowest queuing time. The next step is to find the destination with the highest energy level, out of all the computing resources with the lowest queue time. If the computing resource with the highest energy level and lowest queue time is higher than the receiving UAV's current energy level, then the task will be offloaded to another destination. If not, then the receiving UAV will compute the task locally.

\subsubsection{Deep Q-Learning} \label{sec:DQL}
We used the Deep Q-Learning algorithm presented by Nguyen et al. \cite{Nguyen2022}. 
The Deep Q-Learning algorithm uses the action set defined in (\ref{eq:action}), the state defined in (\ref{eq:state}) and the epsilon-greedy policy. The reward function defined in (\ref{eq:rew1}) consists of two subparts: the reward for choosing an action that does not  increase energy consumption ($\Upsilon^{L}_{j_{a}} - 1$) over threshold $e$, and the reward for selecting a destination that did not lead to a deadline violation $(1-\mathbb{E}(v_{j_{a}})) + \mathcal{V}^{L}_{j_{a}} * \mathbb{E}(v_{j_{a}}))$. Where $\mathcal{V}^{L}_{j_{a}}$, determines the magnitude of the negative reward if a deadline violation occurred when it could have been prevented if the task had been sent to the MEC server ($\mathbb{E}(v_{j_{m}}) = 0$), the task was performed locally ($\mathbb{E}(v_{j_{r}}) = 0$), the task was performed by another UAV ($\exists j'\in(\mathcal{J}/(j_{r}\cup j_{a}))(\mathbb{E}(v_{j'})) = 0$), or a deadline violation was inevitable.
\begin{flalign}
    \label{eq:rew1}
    &\mathbb{R} = (\Upsilon^{L}_{j_{a}} - 1) + (1-\mathbb{E}(v_{j_{a}})) + \mathcal{V}^{L}_{j_{a}} * \mathbb{E}(v_{j_{a}}) \\
    \label{eq:rew2}
	& \Upsilon^{L}_{j_{a}} =
	\begin{cases} 
	        2, & \text{if } \mathbb{E}(\Upsilon^{R}_{j_{a}}) - \max_{j'\in\mathcal{J}}(\mathbb{E}(\Upsilon^{R}_{j'})) \ge -e \\
            0,& \text{if } \mathbb{E}(\Upsilon^{R}_{j_{a}}) - \max_{j'\in\mathcal{J}}(\mathbb{E}(\Upsilon^{R}_{j'})) \le -2e \\
            1, & \text{otherwise,}
	\end{cases}
\end{flalign}
\begin{flalign}
	\label{eq:rew3}
	& \mathcal{V}^{L}_{j_{a}} =
	\begin{cases} 
        -40, & \text{if } \mathbb{E}(v_{j_{m}}) = 0 \\
        -20,& \text{if } \mathbb{E}(v_{j_{r}}) = 0 \\
        -10,& \text{if } \exists j'\in(\mathcal{J}/(j_{r}\cup j_{a}))(\mathbb{E}(v_{j'})) = 0 \\
        -1, & \text{otherwise.}
	\end{cases}
\end{flalign}

\subsubsection{Deep Risk Sensitive (DRS) Reinforcement Learning} \label{sec:DRS}
We extended Pamuklu et al.'s work in \cite{Pamuklu2022} by adding deep learning to their risk-sensitive approach. Instead of using Q-tables to predict the risk and reward Q-values, we used deep neural networks. Deep learning enabled us to extend the state space to include transmission delays between IoT devices and UAVs, UAVs to UAVs, and UAVs to MEC. The state space is defined in (\ref{eq:state}). The action set is defined in (\ref{eq:action}). For each action, the reward and risk is evaluated separately. The reward evaluates how close the agent's action came to achieving objective P2 in \cite{Pamuklu2022}. It is defined as follows,
\begin{flalign}
    \label{eq:rewDRS}
    &\mathbb{R}_{DRS} = - \left[W * (\Upsilon^{L}_{j_{a}}-1) - \frac{1-W}{\Theta^{M}} * \Delta_{j_{a}} \right], 
\end{flalign}

where the $(\Upsilon^{L}_{j_{a}}-1)$ portion is proportional to how the action impacted the battery level of the computing resource that computed the task. $\Delta_{j_{a}}$ is the mean uplink delay of the computing resource $a$ after the task was added to its queue. 

The risk function can be defined as
\begin{flalign}
    \label{eq:riskcost}
    &\mathbb{C} =
	\begin{cases} 
	        -\mathcal{V}^{L}_{j_{a}}, & \text{if } \text{state} \in \mathbb{RS} \\
            -1, & \text{otherwise}
	\end{cases}\\ \notag
	\label{eq:riskstate}
    & \textbf{Where:  } \mathbb{RS} = \{ r | r \subset \mathbb{S}, \text{ and } \mathbb{E}(v_{ja}) \}
    \end{flalign}
It addresses the risk constraint defined in the objective function \cite{Pamuklu2022}. If the agent reaches a risk state, then the agent will be penalized. The magnitude of that penalization is determined by $\mathcal{V}^{L}_{j_{a}}$. If a deadline could have been avoided, then the magnitude of $\mathcal{V}^{L}_{j_{a}}$ is high, otherwise, it is low.

\begin{figure}
\centering
\includegraphics[width=0.442\textwidth]{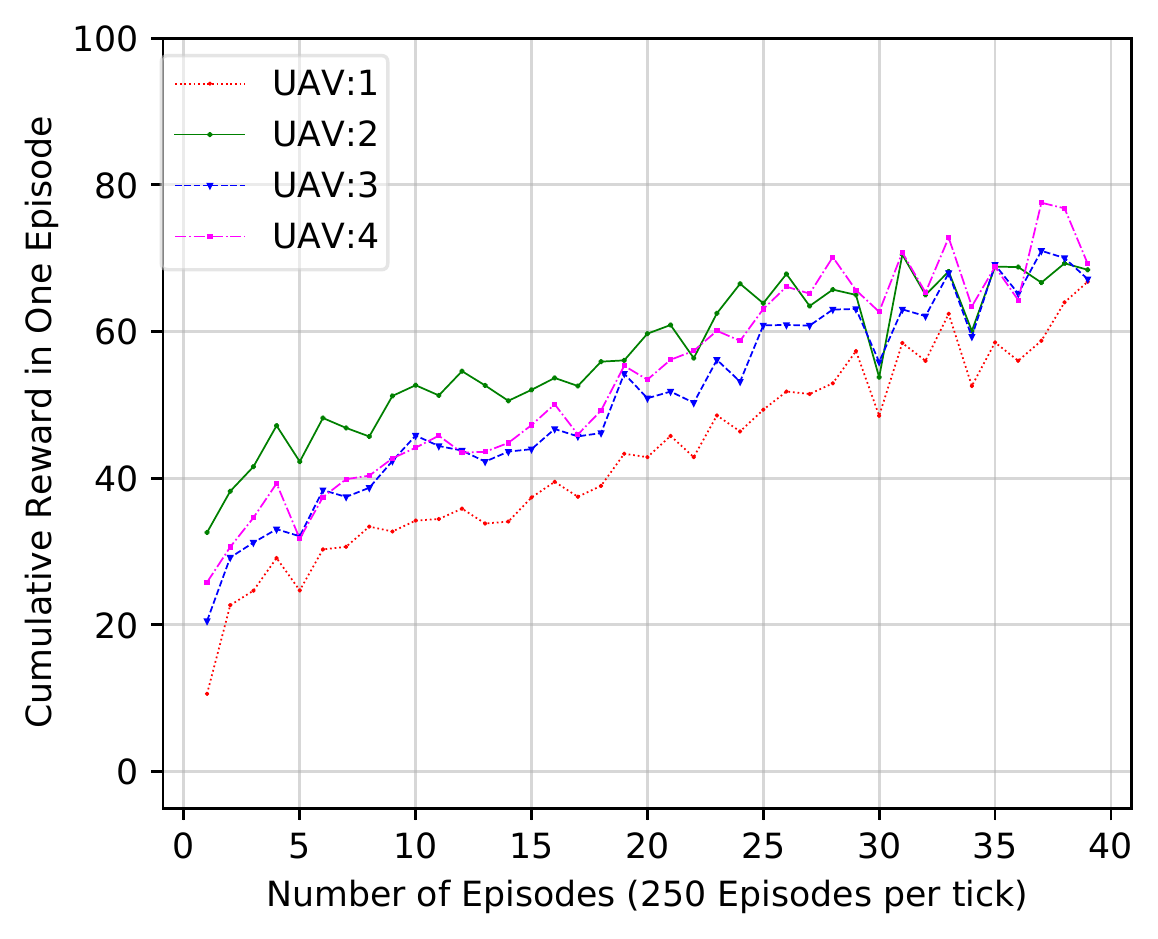}
\caption{\label{fig:CVARconv} Reward convergence for RQ method.}
\end{figure}

\begin{figure}
\centering
\includegraphics[width=0.48\textwidth]
{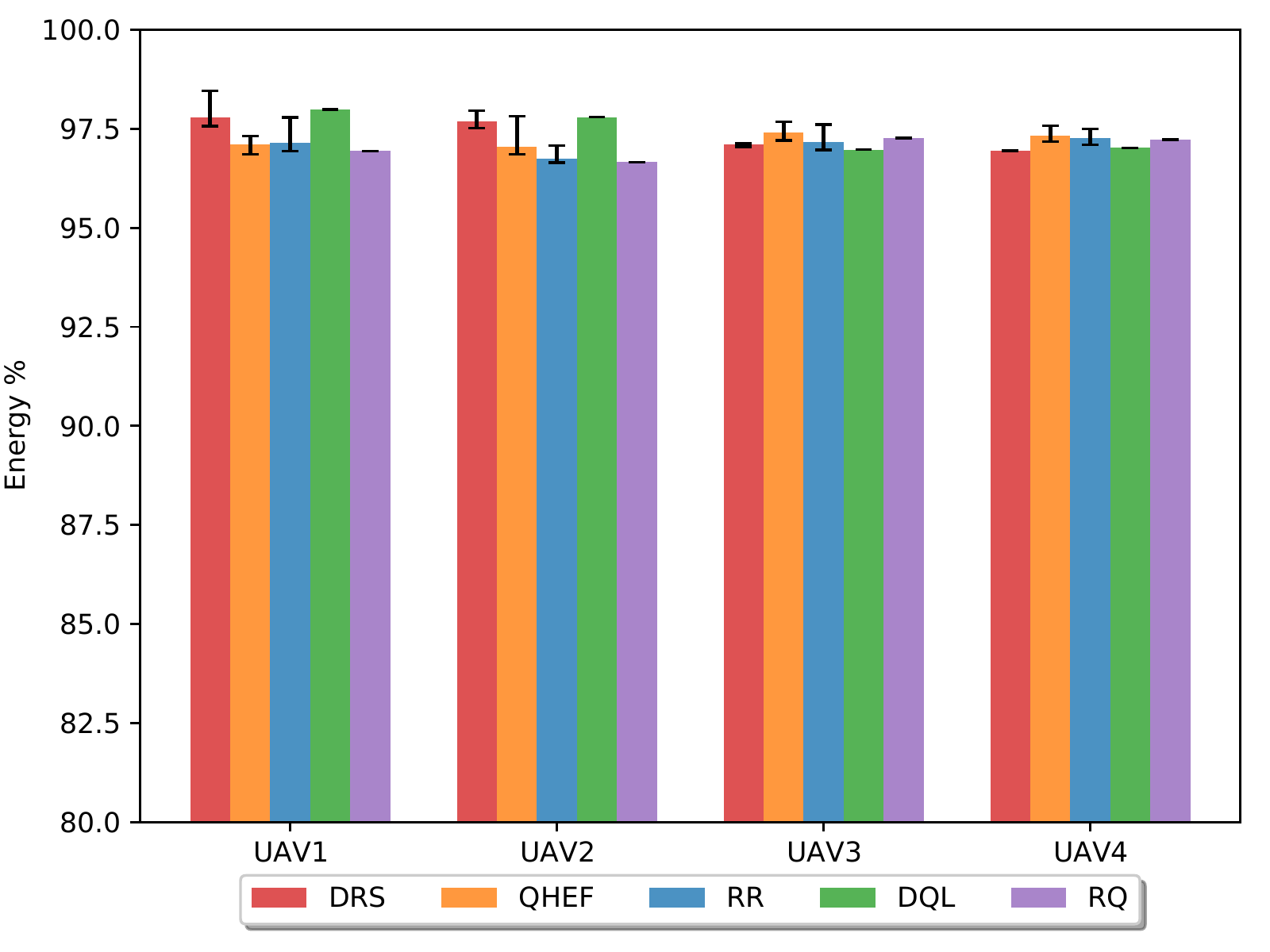}
\caption{\label{fig:CVARBattery} Remaining battery levels for all UAVs across the network.}
\end{figure}

\section{Performance Evaluation} 
In order to simulate our MEC-aided UAV smart farm network, we used  Simu5G \cite{NardiniOmnet}. It is a 5G network simulation library that runs on a network simulator platform called Omnet++ \cite{omnetpp}. Our simulation network consisted of sixteen IoT devices ($X=16$), four UAVs ($J=4$), and one MEC server ($L=1$). The IoT devices generated three types of tasks ($K=3$), fire detection, pest detection, and growth monitoring. Table \ref{tab:SimParams} shows the parameters that we used for each task during our simulation. The tasks' interarrival time (1/$\lambda$) followed an exponential distribution. Each task had a unique deadline ($\gamma^{D}_{\task}$), and the deadlines for all fire detection, pest detection and growth monitoring tasks were 1s, 2s, and 15s respectively. The MEC processing times for each task ($\gamma^{P}_{\task}$ (MEC)) were lower than the UAVs' processing times ($\gamma^{P}_{\task}$ (UAV)) because we assumed that the MEC server had more computing resources and therefore will compute the tasks faster. 

For the energy consumption, the following parameters were used with (\ref{eq:energyeq}), UAVs with index 0 and 1 had full battery capacity ($\Upsilon^{B}_{j'}$) of 570 watt-hours, meanwhile, UAVs with index 2 and 3's full battery capacity was 627 watt-hours. Hovering ($\Upsilon^{H}_{j'}$) consumed 211 watt-hours. The antenna ($\Upsilon^{A}_{j'}$) needed 17 watt-hours. In order to operate the CPU\footnote{Idle and running CPU energy consumptions are selected based on limited simulation time. Thus, UAVs that will operate for ten hours can be simulated, and the difference in energy performance between the methods can be tested.} we needed 4320 watt-hours if the CPU is idle ($\Upsilon^{I}_{j'}$) and 12960 watt-hours if the CPU is computing a task ($\Upsilon^{C}_{j'}$). 

We used the following reinforcement learning parameters, a learning rate of 0.05 and a discount factor of 0.85. Our simulation time was 5s ($\mathcal{T}$ = 5). We trained our model using a dataset of 100 recorded simulations. In terms of the cost function parameters for our Risk Quantifying method, we used an energy risk growth value of 2 ($g=2$), fire task end-to-end delay risk growth value of 8 ($s=8$), and other task risk growth value of 1 ($w=1$). For the risk measurement, we used an $\alpha$ value of 2. In RQ's reward function, $\beta = 0.75$.  

\begin{figure}
\centering
\includegraphics[width=0.48\textwidth]{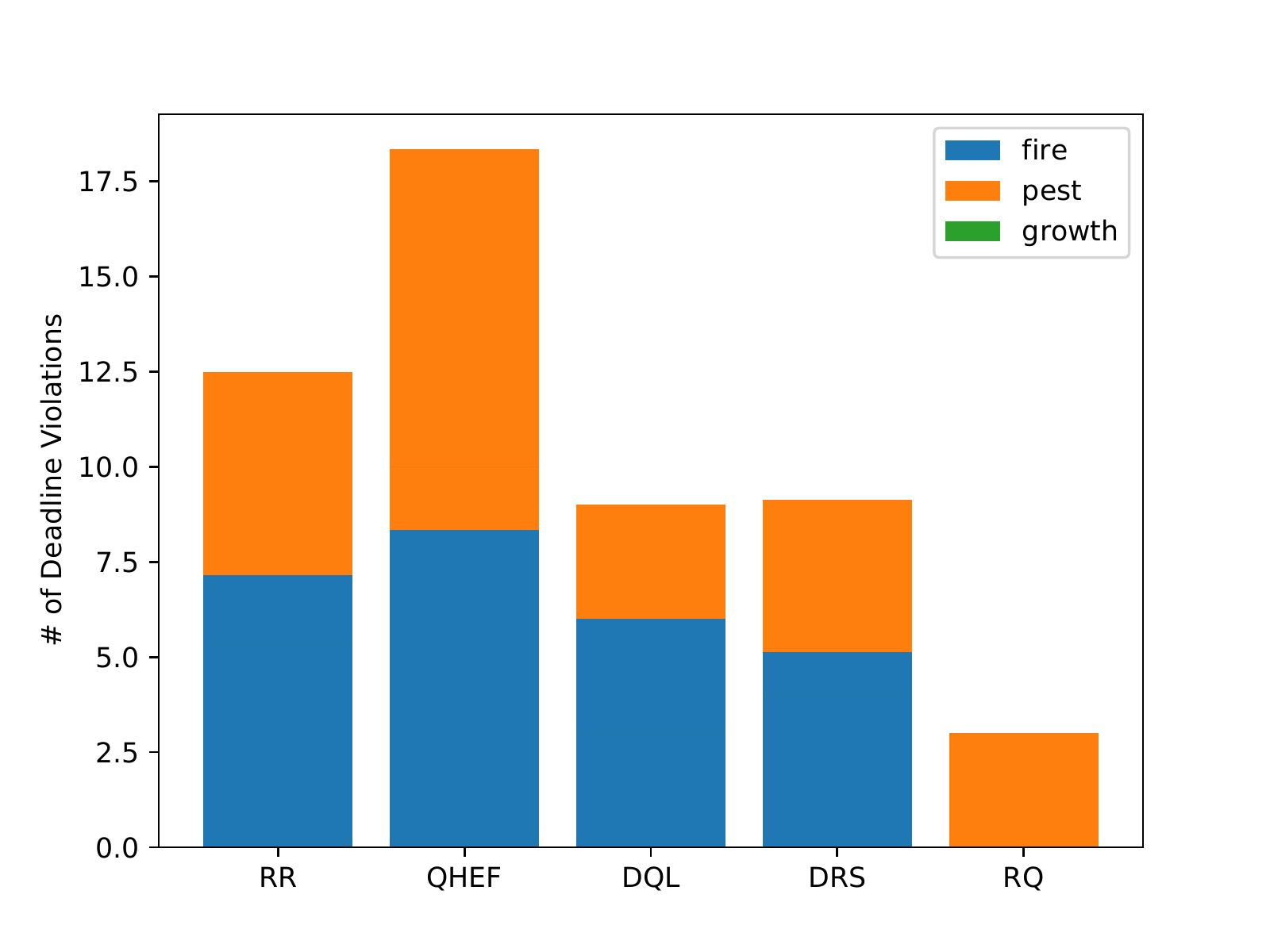}
\caption{\label{fig:CVARTaskDV} Deadline violation distribution per task type.} 
\end{figure}

\par \subsection{Simulation Results} 
Every key performance indicator plot shows the average results over 10 runs with different seeds. 
\begin{table}
    \centering
    \caption{\label{tab:SimParams} Simulation tasks' parameters from \cite{Pamuklu2022}}
    \begin{tabular}{l|c|c|c|c}
    Task Type & $(1/\lambda)$ & $\gamma^{D}_{\task}$ & \begin{tabular}{@{}c@{}}$\gamma^{P}_{\task}$ \\ (UAV)\end{tabular} & \begin{tabular}{@{}c@{}}$\gamma^{P}_{\task}$ \\ (MEC)\end{tabular}\\
    \hline
    Fire detection &  0.25s & 1s & 0.1s & 0.05s\\
    Pest detection &  0.25s & 2s & 0.2s & 0.1s \\
    Growth monitoring & 0.5s & 15s & 1.5s & 0.75s \\
    \end{tabular}
\end{table}

\par \subsubsection{Convergence}
Fig.~\ref{fig:CVARconv} demonstrates the cumulative reward for the proposed RQ method after 10000 episodes. We can see that the reward converges after 8750 episodes. Each UAV is an independently trained agent and has its own deep neural network to predict the most accurate Q-value. Despite being independent of one another, they all have similar performances and converged towards the same value at the end of the training period.

\par \subsubsection{Remaining Energy Level}
The remaining energy level is a KPI used to determine how long the UAV network can remain hovering. A high remaining energy level signifies that the UAV can hover for a longer period of time, whereas a low remaining energy level indicates that the UAV will not hover for a long period of time. In Fig.~\ref{fig:CVARBattery}, QHEF had the highest minimum remaining energy level at 97.04\%, and DQL was not far behind at 96.97\%. RQ's minimum remaining energy level was 96.65\% which is 0.39 lower than QHEF, meaning the difference between their performances is not significant. 

\begin{figure}
\centering
\includegraphics[width=0.48\textwidth]{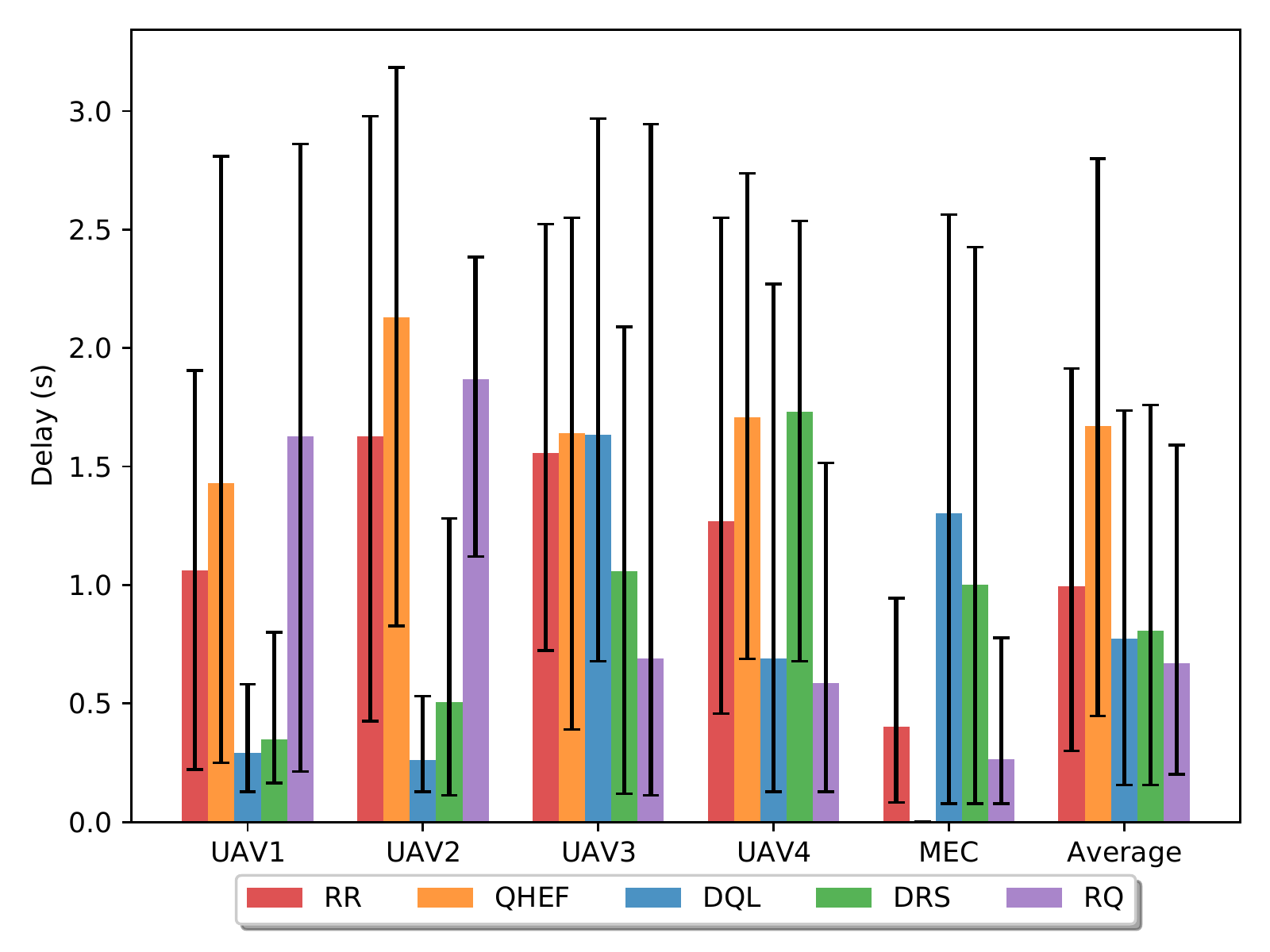}
\caption{\label{fig:CVARuplink} Uplink delay distribution per node.}
\end{figure}

\par \subsubsection{Deadline Violations}
A deadline violation occurs when the uplink delay exceeds the task's deadline. The RQ had the lowest percentage of deadline violations. The ML algorithms outperformed all other algorithms because they made deadline violation a focus in their objective functions. In Fig.~\ref{fig:CVARTaskDV}, we can see that the RQ method was able to outperform the deep risk-sensitive method because it was able to eliminate fire task deadline violations. For all the other methods, the fire task made up most of their deadline violations. 

\par \subsubsection{Mean Uplink Delay}
The uplink delay is the sum of all the transmission delays involved with sending the task to the destination computing resource, the queuing time, and CPU processing time. Fig.~\ref{fig:CVARuplink} illustrates the uplink delay for each algorithm across every node in the network. We can see that the RQ method had the lowest average uplink delay at around 0.6 seconds. This is because it has a cost function that considers the end-to-end delay, and the more the uplink delay exceeds the task's deadline, the higher the cost. This would then lead to the agent's risk costs increasing. The ML algorithms outperformed the heuristic algorithms in uplink delay because they could self-adjust their destination-finding rules and adapt to new situations; meanwhile, the RR and QHEF algorithms had fixed rules.  

\section{Conclusion}
In this study, we presented machine learning algorithms with awareness of risk quantification in an IoT-aided smart farm setting. We used two different techniques to identify risk behaviors in order to train the agent to avoid such behaviors that can lead to severe consequences. The first technique was an extension of our previous study, where we had a separate state to identify whether the agent had entered a risky situation. The second technique used cost functions to evaluate the potential damages of each action and used a risk parameter based on CVaR in order to evaluate the agent's decision-making. We compared our techniques with another deep RL technique and two non-machine learning heuristic algorithms and saw that the risk parameter technique (RQ) was able to completely avoid the risky behavior which is exceeding the deadline on a fire recognition task. This, in turn, reduced the total number of deadline violations and the mean uplink delay. In the future, we would like to include multi-UAV trajectory planning into our problem.

\section*{Acknowledgement}
This  work  is  supported  by MITACS Canada Accelerate program in collaboration with Nokia Bell Labs.
\bibliography{tieraml.bib}

\end{document}